\documentclass[%
 reprint,
 amsmath,amssymb,
prl,
]{revtex4-1}

\usepackage{amsmath}
\usepackage{amssymb}
\usepackage{amsfonts}
\usepackage{dsfont}
\usepackage{graphicx}
\usepackage{bm}
\usepackage{color}
\usepackage{appendix}
\usepackage{epsfig}
\usepackage{dcolumn}
\usepackage{hyperref}
\usepackage{url}

\begin{document}

\preprint{APS/123-QED}

\newcommand\GS{{\it{\Gamma}_{\rm{S}}}}
\newcommand\GD{{\it{\Gamma}_{\rm{D}}}}
\newcommand\GDeff{\it{\Gamma}_{\rm{D}}^{\rm eff}}
\newcommand\Gcav{{\it{\Gamma}}_{\rm{cav}}^{\rm dot}}
\newcommand\Vcav{V_{\rm cav}}
\newcommand\Vdot{V_{\rm dot}}
\newcommand\V{V_{\rm SD}}
\newcommand\Vtbs{V_{\rm TBS}}
\newcommand\Vtbd{V_{\rm TBD}}
\newcommand\Vtbc{V_{\rm TBC}}
\newcommand\Gcavlife{{\it{\Gamma}}_{\rm{cav}}}
\newcommand\eV{{\rm eV}}

\title{Transport Spectroscopy of a Spin-Coherent Dot-Cavity System}

\author{C. R{\"{o}}ssler}
 \email{roessler@phys.ethz.ch}
\author{D. Oehri}
\author{O. Zilberberg}
\author{G. Blatter}
\author{M. Karalic}
\author{J. Pijnenburg}
\author{A. Hofmann}
\author{T. Ihn}
\author{K. Ensslin}
\author{C. Reichl}
\author{W. Wegscheider}

\affiliation{Solid State Physics Laboratory, ETH Zurich, 8093 Zurich, Switzerland}%

\date{\today}

\begin{abstract}
Quantum engineering requires controllable artificial systems with quantum
coherence exceeding the device size and operation time.
This can be achieved with geometrically confined low-dimensional electronic
structures embedded within ultraclean materials, with prominent examples
being artificial atoms (quantum dots) and quantum corrals (electronic
cavities). Combining the two structures, we implement a mesoscopic coupled
dot-cavity system in a high-mobility two-dimensional electron gas, and
obtain an extended spin singlet state in the regime of strong dot-cavity
coupling. Engineering such extended quantum states presents a viable route for
nonlocal spin coupling that is applicable for quantum information processing.
\end{abstract}

\pacs{Valid PACS appear here}
\maketitle

Quantum physics has profited enormously from combining optical cavities and
atoms into a quantum engineering platform, where atoms mediate interactions
among photons and photons communicate information between
atoms~\cite{haroche_nobel_2013}.  In mesoscopic physics, the
constituents of the optical success story have been independently realized:
(i) prototypes of electronic cavities were implemented through structuring of
a two-dimensional electron gas (2DEG) \cite{katine_point_1997,
hersch_diffractive_1999}, yielding extended fermionic modes akin to quantum
corrals on metal surfaces \cite{crommie_imaging_1993, manoharan_quantum_2000},
and (ii) artificial atoms in the form of quantum dots have been
studied, unraveling numerous interesting phenomena such as the Coulomb
blockade \cite{kouwenhoven_few-electron_2001} and the Kondo effect
\cite{goldhaber-gordon_kondo_1998, kouwenhoven_revival_2001}. The combination
of several dots into controlled quantum bits (qubits) has been
demonstrated~\cite{hayashi_coherent_2003, petta_coherent_2005}, thus promoting
the next challenge of introducing coherent coupling between distant qubits
without relying on nearest-neighbor exchange. Analogously to cavity quantum
optics, such coherent coupling could be provided by a suitably engineered
cavity mode.
\begin{figure}[t!]
\includegraphics[scale=0.9]{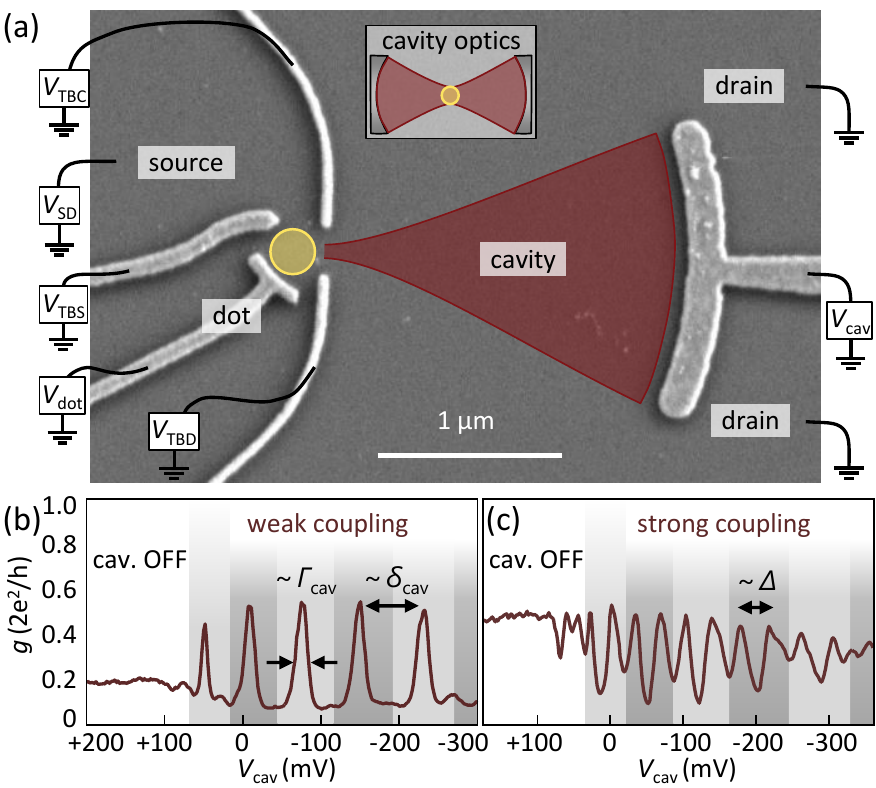}
\caption{\label{fig:fig1}
(a) Scanning electron micrograph of the dot-cavity device. Schottky
electrodes (bright) define an electronic cavity (red overlay) focused onto the tunnel barrier of a
dot (yellow circle overlay). In a typical cavity quantum optics experiment
(inset), a confocal cavity is coupled to an atom.
Our setup resembles a hemispherical cavity where the cavity is focused onto an atom embedded in a plane mirror.
Gate voltages $\Vdot$ and $\Vcav$ change the energies and occupancies of the dot and cavity; $\Vtbs$ and $\Vtbc$ tune the
coupling $\GS$ of the dot to source lead, whereas $\Vtbd$ and $\Vtbs$ tune the couplings
$\GD$ and ${\it\Omega}$ of the dot to both drain and cavity modes simultaneously.
We report measured differential conductance $g=dI/d\V$ of the
dot-cavity system obtained at small bias $\V=-10\,\rm{\mu V}$ in two coupling
configurations $\GD \ll \GS$ (b) and $\GD\gtrsim \GS$ (c). The dot is tuned to
the Coulomb valley of the third charge state. As long as the cavity
is not defined ($\Vcav > 50\,{\rm mV}$), constant transport is due to the
Kondo effect. As the cavity voltage $\Vcav$ depletes the 2DEG, a mirror forms
that focuses quantized modes onto the tunnel barrier of the dot.
(b) Tuning $\Vcav$, the cavity modes are pushed through the chemical potential,
causing peaks of increased Kondo conductance. These peaks are separated
by a cavity-mode spacing $\delta_{\rm cav}\approx 220\,\mu\eV$
and have a width ${\it \Gamma}_{\rm cav}\approx 40 \,\mu\eV$.
(c) Increasing $\GD$, signatures of strong coupling develop when the coupling
${\it\Omega}$ increases beyond the decay rate ${\it \Gamma}_{\rm cav}$.
This strong coupling
manifests in the appearance of split peaks with a gap ${\it\Delta}\approx 80\,\mu\eV$,
indicating the formation of a spin singlet on the dot-cavity hybrid that competes with and
suppresses the Kondo effect.
Different gray shades are applied to the background to highlight the effect of each cavity mode crossing the chemical potential. Note that the two measurements (b) and (c) are slightly shifted and stretched in $\Vcav$ to correct for the different settings of $\Vtbs$ and $\Vtbd$.
}
\end{figure}
Here, we report on transport spectroscopy and control of a coherently coupled
mesoscopic dot-cavity system, where the cavity is embedded in the drain of
the dot. In the weak-coupling regime, we find signatures of standard dot
transport enhanced by a tunnel coupling to drain which is modulated by electrostatic
control of the cavity. In stark contrast, the strong coupling regime exhibits
the formation of a dot-cavity singlet state testifying to the coherent
dynamics of the hybrid system. This singlet competes with and blocks the
formation of a Kondo resonance, signaling the presence of a many-body quantum
phase transition.  Conjoining our measurements with the large spatial scale of
the cavity modes alludes to the dot-cavity hybrid being an original
realization of a Kondo box setup~\cite{thimm_kondo_1999,
dias_da_silva_zero-field_2006, dias_da_silva_spin-polarized_2013}.
Focusing on applications, it constitutes a tunable and purely electrical
component that may serve as a
quantum bus for coherently coupling spatially separated qubits.

The electronic dot-cavity device is shown in Fig.~\ref{fig:fig1}(a).
It resides within a 2DEG, $90\,\rm{nm}$ underneath the surface of a GaAs/AlGaAs
heterostructure. Applying negative voltages to Schottky top gates depletes the
underlying 2DEG and defines a standard dot tunnel coupled to source and drain
leads. An additional arc-shaped gate is situated $\sim 2\,{\rm \mu m}$ away
from the dot with its focal point at the tunnel barrier of the dot. Thus,
applying a voltage $\Vcav$ generates an electronic mirror that confines
quantized ballistic cavity modes with increased weight at the tunnel
barrier~\cite{hersch_diffractive_1999}.  Notably, we designed a cavity gate
with a relatively small opening angle of $45^\circ$ in order to confine only
fundamental one-dimensional modes, i.e., high angular-momentum modes leak
out into the drain. This guarantees unique identification and addressability of individual
cavity modes.

We perform equilibrium transport spectroscopy of the dot-cavity system at an
electronic temperature $T_{\rm{el}}<20\,\rm{mK}$~\cite{baer_experimental_2014,
rossler_spectroscopy_2014}.  The dot is tuned to the Coulomb blockade of
the third electron-charge state. In this configuration, linear transport is
dominated by the Kondo effect arising due to screening of the unpaired
electron spin on the dot by the lead electrons~\cite{hewson_kondo_1997,
glazman_resonant_1988, goldhaber-gordon_kondo_1998}.  In Figs.~\ref{fig:fig1}(b)
and \ref{fig:fig1}(c), we report on two measurements of the differential
conductance $g=dI/d\V$  as a function of $\Vcav$ with weak coupling and strong
coupling between the dot and the drain-cavity system, respectively.  In both configurations,
we observe a constant Kondo conductance as long as the cavity gate does not
deplete the 2DEG. Once the 2DEG below the cavity gate is depleted, an
electronic cavity is formed with its states filled up to the chemical
potential. Taking the lithographically defined length of $L_{\rm
cav}=1.9\,{\rm \mu m}$ and the Fermi wavelength $\lambda_{\rm F}\approx
53\,{\rm nm}$, we estimate that upon formation $n_{\rm cav}\approx 2L_{\rm
cav}/\lambda_{\rm F}\approx 70$ states are filled.  Applying increasingly
negative $\Vcav$, the cavity becomes shorter and its states rise in energy. As
these states are focused on the tunnel barrier of the dot, they effectively enhance the tunnel
coupling $\GD$ when passing through the chemical potential of the leads. This
is seen in Fig.~\ref{fig:fig1}(b) as a series of peaks with enhanced Kondo
transport. In the strong coupling regime (see Fig.~\ref{fig:fig1}(c)), we
observe that these peaks are split and Kondo transport is quenched.
We interpret this reduction as the result of a spin singlet
formation within the dot-cavity hybrid that competes with the Kondo screening.
The inferred extent of this coherent singlet state over the entire dot-cavity
system is the main result of this Letter.

\begin{figure}[t]
\includegraphics[scale=0.95]{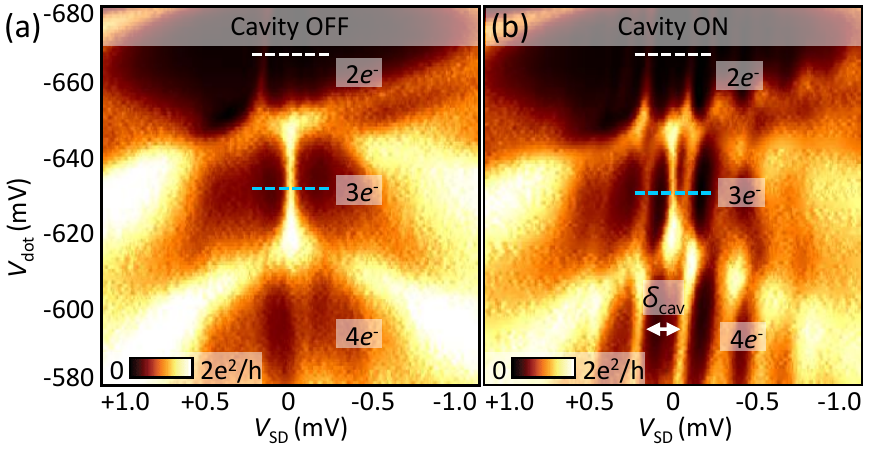}
\caption{\label{fig:fig2}
(a) Differential conductance of the dot plotted as a function of $\V$ and
$\Vdot$ with the cavity switched off ($\Vcav=+200\,\rm{mV}$) and symmetric tunnel
coupling $\GD\approx \GS$. Characteristic (dark) regions of the Coulomb blockade
are intersected by a pronounced zero bias Kondo resonance at odd dot
occupation ($N=3\,e^-$).  (b) Same experimental parameters, but with
cavity modes present ($\Vcav=-200\,\rm{mV}$) in the drain of the dot.
Additional resonance lines spaced by $\delta_{\rm cav}\approx~220\,\mu \eV$
arise due to the cavity modes modifying the dot transport. Horizontal dashed
lines at fixed $\Vdot$ indicate the parameter settings used in Fig.~\ref{fig:fig4}(b) (blue line) and
Fig.~\ref{fig:fig4}(e) (white line). Note that increasing $\Vdot$ also increases
the tunnel couplings, resulting in more pronounced transport signatures in the
bottom of the figures.
}
\end{figure}

To better illuminate the impact of the cavity on standard dot transport, we
compare finite-bias measurements of $g=dI/d\V$ in the absence
[Fig.~\ref{fig:fig2}(a)] and presence [Fig.~\ref{fig:fig2}(b)] of the cavity as a
function of a plunger gate voltage $\Vdot$ that shifts the dot levels.
The drain tunnel barrier is tuned to strong coupling with $\GS\approx\GD$.
Without the cavity ($\Vcav=+200\,\rm{mV}$), the typical pattern of a
few-electron dot exhibits regions of suppressed conductance (Coulomb diamonds),
resulting from blockade of the respective charge states $N=2e^-,3e^-,4\,e^-$. As the
occupation of the dot increases, its orbital wave function is spatially more
extended, thus enabling stronger tunnel coupling to the
leads~\cite{kouwenhoven_few-electron_2001}.  A pronounced zero bias
Kondo resonance appears in the Coulomb valley of $N=3\,e^-$.
With the cavity switched on ($\Vcav=-200\,\rm{mV}$), we observe additional features
in Fig.~\ref{fig:fig2}(b), i.e., new lines of increased conductance that pass through the
regions of the Coulomb blockade, as well as
pronounced modulation of the direct transport at the boundaries of the Coulomb
diamond.  Tuning $\Vcav$ (not shown) sweeps these novel features horizontally
through the Kondo resonance leading to its controlled modulation seen in
Fig.~\ref{fig:fig1}(c).  Notably, Figs.~\ref{fig:fig2}(a) and b allow us to
characterize the system, i.e., determine the tunnel coupling constants
$\GS\approx \GD\approx 87\,\mu \eV$ for this configuration, its charging energy
$U\approx700\,\mu \eV$, and the cavity level spacing $\delta_{\rm
cav}\approx 220\,\mu \eV$.
\begin{figure*}
\includegraphics[scale=0.95]{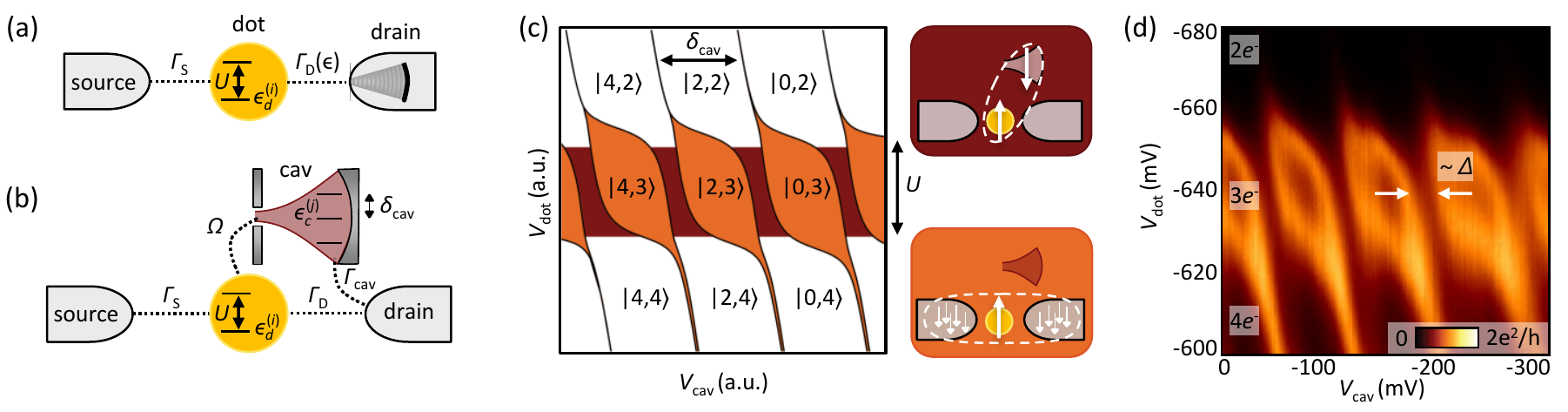}
\caption{\label{fig:fig3}
(a) A sketch of an Anderson model depicting a dot with energies
$\epsilon_{d}^{\scriptscriptstyle (i)}$, on-site interaction $U$, constant
coupling $\GS$ to the source, and energy-dependent coupling $\GD(\epsilon)$ to
the drain due to focused cavity modes~\cite{oehri_prep_2015}. In this model the effect of the cavity mirror on the dot is encoded as a structured tunneling amplitude, $\GD(\epsilon)$, into the drain lead.  (b) A sketch of a coherent dot-cavity model where the
$\GD(\epsilon)$  of the previous model is replaced by a noninteracting cavity
with a  discrete spectrum $\epsilon_{c}^{\scriptscriptstyle (j)}$, effectively coupled to the dot
with ${\it\Omega}$ and to the drain lead by $\Gcavlife$. (c) Calculated ground state
map $|N_{\rm cav},N_{\rm dot}\rangle$ of the dot-cavity hybrid as a function
of $\Vcav$ and $\Vdot$ with (relative) occupation numbers $N_{\rm dot}$ ($N_{\rm cav}$)
determining the ground state of the dot (cavity) for ${\it \delta}_{\rm cav}/U=0.5$ and
${\it\Omega}/U=0.1$. Dark red shading marks the regimes of spin singlet
formation on the dot-cavity hybrid, where the Kondo resonance (orange) is suppressed.
(d) Measurement of the linear conductance (with $\V=-10\,\rm{\mu V}$) in the strong coupling regime.
Lines of resonant transport match the boundaries between ground states in (c).
}
\end{figure*}

The cavity modes in the drain exhibit a remarkable coherence in view of the various relaxation and scattering processes within the drain. These
modes affect the dot transport by modulating the tunneling density of states to the drain. A theoretical model that corresponds to this picture is sketched in Fig.~\ref{fig:fig3}(a). Analyzing the structure of the tunnel coupling to the drain, we are able to distinguish separated coherent cavity modes that are broadened by coupling to a bath~\cite{oehri_prep_2015}. The latter indicates that the cavity modes constitute an additional degree of freedom that is coupled to the dot, as described by the theoretical model sketched in Fig.~\ref{fig:fig3}(b). We assume that charging effects on the cavity are screened due to its large extent and embedding in the drain lead, i.e., we do neither include on-site cavity nor mutual dot-cavity Coulomb interactions in our model.  Analyzing the hybridization of dot and cavity modes by exact diagonalization of their joint Hamiltonian results in molecular states akin to double-dot structures.  Hence, as a function of $\Vdot$ and $\Vcav$, different ``molecular'' ground state occupations occur depending on the dot and cavity level configuration relative to the Fermi energy, see Fig.~\ref{fig:fig3}(c).  Tuning $\Vdot$ changes the dot occupation by a single electron as the dot moves between Coulomb valleys. In contrast, the noninteracting cavity levels can change their occupancy by two electrons as a function of $\Vcav$. Nonetheless, regions of odd cavity occupation manifest themselves as well (dark red in the figure) due to the exchange coupling with the dot whenever the dot occupation is odd. The ground state in this configuration is a spin singlet extending over the entire dot-cavity system.

Measuring the linear conductance as a function of both $\Vcav$ and $\Vdot$
shows remarkable agreement with this picture:
enhanced transport [Fig. \ref{fig:fig3}(d)] occurs at the boundaries between different
ground states, i.e., whenever the molecular occupation changes [Fig. \ref{fig:fig3}(c)].
The molecular- and the standard dot transport (i.e., cotunneling-, resonant-, and Kondo-) are intimately
related: in the cotunneling regime, dominant at even dot occupancy,
the cavity generates enhanced transport lines (as a function of $\Vcav$) whenever its occupancy
changes by two electrons. In the Kondo regime characteristic of odd dot occupancy,
the enhancement is split into two lines when the
cavity occupation is consecutively filled by single electrons. In between these lines, Kondo transport
is quenched due to the formation of the dot-cavity spin singlet.
Using the model depicted in \ref{fig:fig3}(b), we predict this gap to be
${\it\Delta}\approx 12 {\it\Omega}^2/U$.  This explains the gap formation discussed in
Fig.~\ref{fig:fig1}(c).  Finally, the resonant transport (Coulomb peaks) occurring when the dot
changes its occupancy is similarly split by the strong coupling to the cavity
modes. The compliance between the transport measurements and the coherent
model in Fig.~\ref{fig:fig3}(b) suggests that the dot-cavity setup constitutes
a novel realization of the Kondo box scheme~\cite{thimm_kondo_1999}, where a
magnetic impurity (the dot spin) is screened by a metallic grain with a
discrete spectrum (the cavity).

\begin{figure*}
\includegraphics[scale=0.95]{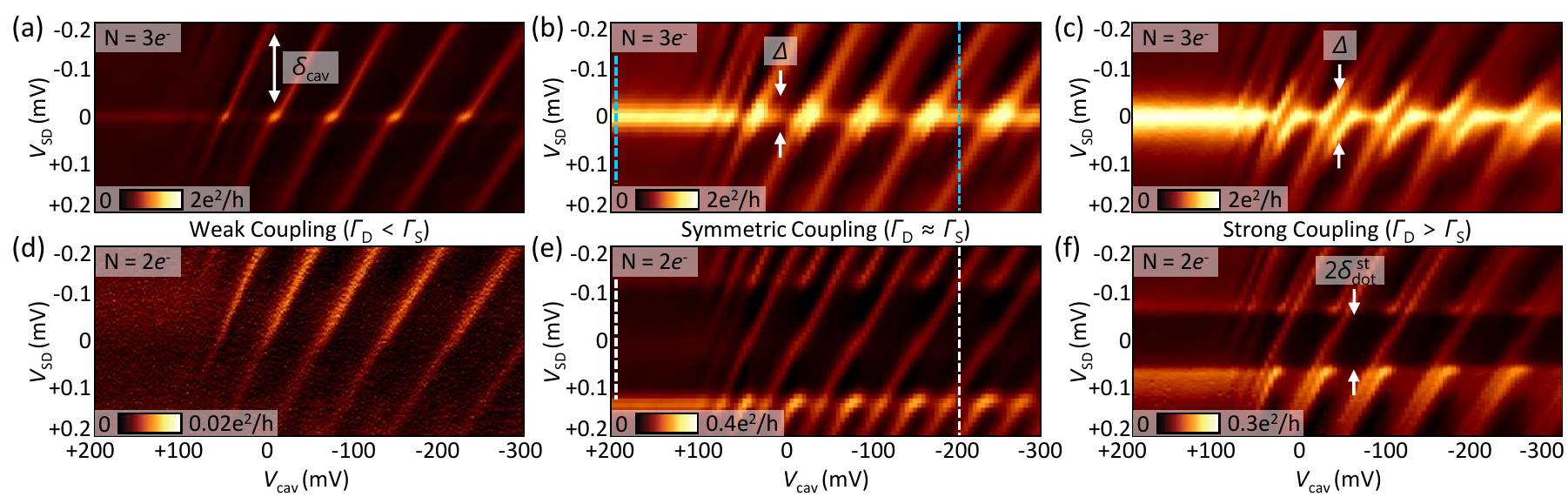}
\caption{\label{fig:fig4}
Finite bias differential conductance of the dot-cavity system (a to c) and control experiment
with even dot occupancy (d to f). (a) In the weak dot-cavity coupling regime,
the odd occupancy of the dot ($N=3\,e^-$) gives rise to a zero bias Kondo
resonance (faint horizontal line) and finite-bias cotunneling signatures
(diagonal lines) due the presence of cavity modes. For increasing coupling
(b), the hybridization between dot and cavity leads to a splitting of Kondo-
and cotunneling features, increasing upon stronger coupling (c). (d)
to (f): Control experiment with $N=2\,e^-$ in the dot, where the Kondo effect
is absent and the cavity resonances do not exhibit an energy gap when the
dot-cavity coupling is increased.
Notably, replicas of the cavity-enhanced cotunneling lines
appear at $\V$ larger than (or equal to) the dot's singlet--triplet energy gap ${\it \delta}_{\rm dot}^{\rm st}$, signaling that the strong
hybridization of dot-cavity alters also the inelastic cotunneling through the triplet states of the dot.
Note the adjusted color scales in (d), (e), and (f) to improve the visibility of the
cotunneling currents. Dashed lines in (b) and (e) indicate the correspondence to Figs.~\ref{fig:fig2}(a) and (b).
}
\end{figure*}

Equipped with this knowledge, we revisit the results of Fig.~\ref{fig:fig1}
and extend them to include out-of-equilibrium signatures as a function of $\V$.
Figures~\ref{fig:fig4}(a)-(c) present a controlled cross-over from weak to
strong coupling between the dot and the drain-cavity system.
This is obtained by a stepwise increase of ${\it\Omega}$ and $\GD$ alongside a decrease of $\GS$.
The zero bias Kondo resonance is modulated [see (a)] as the cavity modes cross the
leads' Fermi level. At finite bias, cotunneling transport is enhanced whenever
the cavity modes are aligned with the source lead. The enhanced cotunneling
lines connect at zero bias to Kondo transport that benefits from the same enhanced
coupling. At stronger coupling (see Figs.~\ref{fig:fig4}(b)-(c)) the transport exhibits a splitting
of both cotunneling and Kondo signatures due to the dot-cavity singlet formation.
We measure the spin singlet gaps in Figs.~\ref{fig:fig4}(b) and (c) to be ${\it\Delta}=32\,\mu\eV$
and ${\it\Delta}=80\,\mu\eV$ corresponding to an effective dot-cavity coupling
${\it\Omega}\approx 43\,\mu\eV$ and ${\it\Omega}\approx 70\,\mu\eV$, respectively.
Comparing these values of ${\it\Omega}$ to the estimated cavity lifetime
${\it \Gamma}_{\rm cav} \approx 40\,\mu\eV$ allows us to establish the regime of strong
coupling through the condition ${\it\Omega}\gtrsim{\it \Gamma}_{\rm cav}$. In
Figs.~\ref{fig:fig4}(d)-(f), we report an experiment repeating the measurements in
Figs.~\ref{fig:fig4}(a)-(c) but with the dot tuned to reside in the Coulomb valley of two
electrons. Because of the absence of an unpaired spin in the dot, both Kondo screening
and dot-cavity singlet formation signatures are removed, i.e., neither zero bias
conductance resonance nor splittings of cavity-enhanced cotunneling lines show
up. Besides serving as a control experiment, Figs.~\ref{fig:fig4}(d)-(f) hold an
additional signature of the strong hybridization between dot and cavity. The dot
transport in the two-electron state exhibits an additional signature at finite
bias voltage due to inelastic cotunneling via the dot's two-electron triplet excited
state~\cite{kyriakidis_voltage-tunable_2002,meunier_experimental_2007}. The
onset of these additional cotunneling processes appears as
enhanced horizontal lines at finite bias voltage. The cavity modulates these
inelastic cotunneling features in the same way as the standard cotunneling
features, such that one cavity mode gives rise to two additional parallel lines
at opposite $\V$, offset by twice the singlet--triplet splitting ${\it \delta}_{\rm dot}^{\,\rm st}$
of the dot, in agreement with the model defined in Fig.~\ref{fig:fig3}(b).

The observed transport signatures through our dot-cavity setup point to
coherent coupling between its constituents which manifests as a spin singlet
extended over the entire device. This has important ramifications for both
fundamental questions in many-body physics, as well as quantum engineering
prospects.  The former touches upon  the study of many-body quantum phase
transitions~\cite{hewson_kondo_1997}. The emerging dot-cavity model is
similar in structure to the Kondo box problem that predicts  splitting of the
Kondo peaks with anomalous scaling~\cite{thimm_kondo_1999}. Thus, our setup
conforms to a controllable experimental realization of such a mechanism
competing with the standard Kondo resonance. This places our results alongside
notable mechanisms that compete with the standard Kondo effect, such as Ruderman-Kittel-Kasuya-Yosida interactions~\cite{craig_tunable_2004}, two-channel
Kondo~\cite{potok_observation_2007}, and singlet--triplet switching on a
molecule~\cite{roch_quantum_2008}. In the future, it will be interesting to study the temperature and magnetic field scaling of the Kondo splitting \cite{thimm_kondo_1999, dias_da_silva_zero-field_2006, dias_da_silva_spin-polarized_2013}. On the quantum engineering front, the
cavity modes can be shaped to connect distant dots~\cite{mehl_two-qubit_2014,srinivasa_tunable_2015}.  As the spin coherence is
conserved, and spin-polarized currents may be
produced~\cite{dias_da_silva_spin-polarized_2013}, this platform holds great
promise for quantum information processing applications.

\section*{Acknowledgments}
We thank M. Kroner, A. Wallraff, G. Scalari, E. van Nieuwenburg, and R. Chitra
for fruitful discussions. We acknowledge the support of the ETH FIRST
laboratory and financial support of the Swiss National Science Foundation
through the NCCR Quantum Science and Technology grant.

\bibliographystyle{apsrev}
\bibliography{Bibliothek}

\end{document}